\documentclass[onecolumn, floatfix, preprintnumbers, eqsecnum, letterpaper, superscriptaddress, nofootinbib]{revtex4}
\usepackage{graphicx}
\usepackage{microtype}
\usepackage{amsmath}
\usepackage{amssymb}
\usepackage{subfigure}
\usepackage{hyperref}
\usepackage{url}
\usepackage{xcolor}
\usepackage{color}
\usepackage{mathrsfs}
\usepackage{calrsfs}
\usepackage{amsfonts}
\usepackage{eufrak}
\usepackage{tabularx}
\usepackage{eucal}
\usepackage{latexsym}
\usepackage{ragged2e}
\usepackage{epsfig}
\usepackage{textcomp}

\usepackage{caption}
\usepackage{subfigure}

\DeclareCaptionJustification{justified}{\leftskip=0pt \rightskip=0pt \parfillskip=0pt plus 1fil}
\definecolor{vividviolet}{rgb}{0.62, 0.0, 1.0}
\definecolor{amaranth}{rgb}{0.9, 0.17, 0.31}
\definecolor{palatinateblue}{rgb}{0.15, 0.23, 0.89}
\definecolor{brightpink}{rgb}{1.0, 0.0, 0.5}
\definecolor{cornflowerblue}{rgb}{0.39, 0.58, 0.93}
\definecolor{deepcarminepink}{rgb}{0.94, 0.19, 0.22}
\definecolor{radicalred}{rgb}{1.0, 0.21, 0.37}

\hypersetup{ linktoc=all,
    colorlinks, linkcolor={palatinateblue},
    citecolor={brightpink}, urlcolor={amaranth}
}

\graphicspath{{Images/}}

\graphicspath{{Images/}}

\def\@fnsymbol#1{\ensuremath{\ifcase#1\or \ddagger \or  $\textleaf$  \or \dagger
\else\@ctrerr\fi}}%

\makeatother

\def\sideremark#1{\ifvmode\leavevmode\fi\vadjust{\vbox to0pt{\vss
 \hbox to 0pt{\hskip\hsize\hskip1em
 \vbox{\hsize1.3cm\tiny\raggedright\pretolerance10000
 \noindent #1\hfill}\hss}\vbox to8pt{\vfil}\vss}}}%

\def\beq{\begin{equation}}
\def\eeq{\end{equation}}

\newcommand{\od}{\mathrm{d}}

\begin{document}

\title{Equation of the Perfect Fluid in the FRW Universe}

\author{Shi-Bei Kong}
\email{shibeikong@ecut.edu.cn}
\affiliation{School of Science, East China University of Technology, Nanchang 330013, Jiangxi, China}

\author{Ying Wang}
\email{2024214413@ecut.edu.cn}
\affiliation{School of Science, East China University of Technology, Nanchang 330013, Jiangxi, China}

\author{Yu-Ke Wang}
\email{2024216357@ecut.edu.cn}
\affiliation{School of Nuclear Science and Engineering, East China University of Technology, Nanchang 330013, Jiangxi, China}

\author{Xiao-Min Wu}
\email{2024214360@ecut.edu.cn}
\affiliation{School of Chemistry and Materials Science, East China University of Technology, Nanchang 330013, Jiangxi, China}

\begin{abstract}

In this paper, we study the equation of state and its properties of the perfect fluid in the $D$-dimensional FRW universe under Einstein gravity, Gauss-Bonnet gravity and Lovelock gravity. In Einstein gravity, we get the equation of state and find that it has no critical point in the $P$-$V$ diagram, but its isothermal lines have minima in the $4$-dimensional case and are always negative in higher dimensions. In Gauss-Bonnet gravity,
we get the equation of state and find that it has a critical point in the $5,6,7,8$-dimensional cases with phase transitions above the critical temperature. In Lovelock gravity, we get the equation of state and conditions of the critical points. Our work shows that both the theories of gravity and the dimensions
of the FRW universe affect the existence of the critical point of the perfect fluid. Interestingly, if the critical point exists, phase transition always occures above the critical temperature.

\end{abstract}

\maketitle

\section{Introduction}

In recent years, one of the most interesting discoveries in black hole thermodynamics is the van der Waals(vdW)-like 
equation of state\cite{Kubiznak:2012wp,Kubiznak:2016qmn}. For the anti-de Sitter(AdS) or de Sitter(dS) black holes 
in the extended phase space\cite{Gunasekaran:2012dq,Cai:2013qga,Xu:2015rfa,Wei:2020poh},
the cosmological constant can be treated as a variable, so its Hawking temperature\cite{Hawking:1975vcx} $T$ can be regarded as a function of the horizon radius $r_h$ and cosmological constant $\Lambda$ as well as some other parameters, i.e. $T=T(r_h,\Lambda,...)$. What's more, the cosmological constant
can be interpreted as the thermodynamic pressure of the black hole, i.e.\cite{Kubiznak:2012wp,Kubiznak:2016qmn}
\begin{alignat}{1}
P\equiv-\frac{\Lambda}{8\pi},
\end{alignat}
and the horizon radius is related to the thermodynamic volume, i.e. $V=V(r_h)$,
e.g. $V=4\pi r_h^3/3$ for the 4-dimensional spherically symmetric black holes.
Therefore, the Hawking temperature $T$ can be regarded as a function of the thermodynamic volume $V$ and the thermodynamic pressure $P$
as well as other variables, i.e. $T=T(V,P,...)$, which can also be written as $P=P(V,T,...)$ and is a vdW-like equation of state.
It has been found that for many black holes in AdS and dS spacetime, their equations of state have vdW-like critical points in the $P$-$V$ diagram
that satisfy
\begin{alignat}{1}
\left(\frac{\partial P}{\partial V}\right)_{T}=0, \quad \left(\frac{\partial^2 P}{\partial V^2}\right)_{T}=0, \label{CC}
\end{alignat}
and there are phase transitions below the critical temperature\footnote{The first example is the charged black hole in anti-de Sitter
spacetime\cite{Kubiznak:2012wp}.}. Below the critical temperature, there are two phases or states-the small state and the large state
similar to the liquid state and gas state of the vdW system, and they can co-exist at certain temperature and pressure
with the same free energy. Apart from the co-existence states, the free energy of the two states is different, 
and the state with a lower free energy is thermodynamically stabler, so there will be a phase transition between the two states 
as pressure or temperature moves across the co-existence value. This is a first order phase transition as the first order derivative of the free energy 
is not continuous at this point although the free energy itself is continuous. Above the critical temperature, the two phases cannot be distinguished,
i.e. there is only one phase, so the free energy and its arbitrary order derivatives are always continuous, thus there is no phase transition. 
Interestingly, at the critical point, the free energy and its first order derivative are continuous 
but its second order derivative is not continuous, so there is a second order phase transition.
Near the critical point, one can also define critical exponents\cite{Pelissetto:2000ek}, which were found to be the same with those in
the vdW system and the mean field theory in nearly all the cases, so they satisfy the scaling laws. 
Recently, critical exponents beyond mean field theory have been found, which even violate the scaling laws\cite{Hu:2024ldp,Kong:2025lqz}.
People also found that black holes can undergo Joule-Thomson expansion\cite{Okcu:2016tgt}, have reentrant phase transition\cite{Altamirano:2013ane}
and triple point\cite{Altamirano:2013uqa}, etc.

Given the thermodynamic similarities between the F(L)RW(Friedmann-(Lemaître)-Robertson-Walker) universe and black holes, 
one expects the FRW universe to also exhibit a vdW-like 
equation of state\footnote{It should be noted that in the cosmology community, `equation of state' is usually referred to as the relation between
pressure $p$ and energy density $\rho$, i.e. $p=\omega\rho$, which is different from the vdW-like equation. In this paper, we use `equation of state'
to refer to the vdW-like equation as is often done in the study of black hole phase transitions.}.
The thermodynamic pressure of the FRW universe is identified with the 
work density\cite{Hayward:1997jp,Cai:2005ra,Akbar:2006kj,Cai:2006rs,Abdusattar:2021wfv,Kong:2021dqd,Kong:2022xny,Abdusattar:2023hlj,Chu:2025zuz},
\begin{alignat}{1}
W\equiv-\frac{1}{2}h_{ab}T^{ab}=\frac{1}{2}(\rho-p), \quad a, b=0,1,
\end{alignat}
where $\rho$ and $p$ are the energy density and pressure of the perfect fluid respectively. It can be written as an equation of state $P=P(V,T)$, where $V$ is the thermodynamic volume and $T$ is the Hawking temperature of the FRW universe. The equation of state depends on the theory of gravity and the dimension
of spacetime. For the (3+1)-dimensional FRW universe, it has been obtained in Einstein gravity\cite{Abdusattar:2021wfv}, scalar-tensor 
theories\cite{Kong:2021dqd,Abdusattar:2023hlj}, the brane world scenario\cite{Kong:2022xny}, and quasi-topological gravity\cite{Chu:2025zuz}, etc. 
It has also been obtained for the high dimensional FRW universe in Einstein-Gauss-Bonnet gravity\cite{Saavedra:2023lds}\footnote{I thank Haximjan Abdusattar for drawing my attention to this paper.}.
In many cases, the vdW-like phase transitions are found, but they occur above the critical temperature,
which is different from the black hole cases.

In this paper, we are interested in the behavior of the perfect fluid in the FRW universe.
The perfect fluid in the FRW universe also has a pressure as well as an energy density. If the perfect fluid is in equilibrium
with the Hawking radiation from the apparent horizon of the FRW universe, its temperature is also equal to the Hawking temperature.
From the corresponding Friedmann's equations, one can also get the equation of state for the perfect fluid in the FRW universe.
The dimension of spacetime also affects the thermodynamic behaviors of the perfect fluid, so we study the perfect fluid 
in the general $D$-dimensional(or $d+1$ dimensional) FRW universe instead of the $4$-dimensional(or $3+1$ dimensional) FRW universe.
In this paper, we only consider three common tensor theories of gravity, i.e. Einstein gravity, Gauss-Bonnet gravity and Lovelock gravity\cite{Lovelock:1971yv,Deruelle:1989fj,Padmanabhan:2013xyr},
and obtain the corresponding equation of state for the perfect fluid in the $D$-dimensional FRW universe. 
In some cases, the vdW-like phase transitions are also found, which depend on the theory of gravity and dimension of the FRW universe
as expected. Phase transitions of the perfect fluid also occur above the critical temperature.

This paper is organized as follows. In Sec.II, we briefly introduce the $D$-dimensional FRW universe. In Sec.III, we obtain the equation of state for the perfect fluid in the $D$-dimensional FRW universe under Einstein gravity. In Sec.IV, we obtain the equation of state and find critical points for the perfect fluid in the $D$-dimensional FRW universe under Gauss-Bonnet gravity. In Sec.V, we obtain the equation of state and discuss its critical points for the perfect fluid in the $D$-dimensional FRW universe under Lovelock gravity. The last section is conclusions and discussions.
We use natural units $c=\hbar=k_B=G=1$.

\section{A Brief Introduction of the $D$-dimensional FRW Universe}

The metric for the $D$(or $n+1$)-dimensional FRW universe can be written as\cite{Akbar:2006kj,Cai:2006rs}
\begin{alignat}{1}
\od s^2=-\od t^2+a^2(t)\left(\frac{\od r^2}{1-kr^2}+r^2\od\Omega_{n-1}^2\right), \label{metric}
\end{alignat}
where $a(t)$ is the time-dependent scale factor\footnote{In the following, we may use $a$ instead of $a(t)$ for convenience.}, $k$ is the spatial curvature (with $+1,0,-1$ corresponding to spatially closed, flat and open respectively), and $\od\Omega_{n-1}^2$ is the metric of the unit $(n-1)$-dimensional sphere.
One can also  rewrite the metric as
\begin{alignat}{1}
\od s^2=h_{ab}\od x^a\od x^b+R^2\od\Omega_{n-1}^2,
\end{alignat}
where $R:=a(t)r$ is the physical radius and $h_{ab}$ is the $2$-dimensional metric with $a,b=0,1$ and $x^0=t, x^1=r$.

The apparent horizon of the FRW universe is defined by $h^{ab}\nabla_a R\nabla_b R=0$, and its solution is
\begin{alignat}{1}
R_A=\frac{1}{\sqrt{H^2+\frac{k}{a^2}}}. \label{AH}
\end{alignat}
The Hawking temperature of the FRW universe is
\begin{alignat}{1}
T=\frac{1}{2\pi R_A}\left(1-\frac{\dot{R}_A}{2HR_A}\right). \label{HT}
\end{alignat}

For the FRW universe, its matter is regarded as a perfect fluid with energy-momentum tensor
\begin{alignat}{1}
T_{\mu\nu}=(\rho+p)U_{\mu}U_{\nu}+pg_{\mu\nu}, \label{pf}
\end{alignat}
where $\rho$ is its energy density, $p$ is its pressure and $U^{\mu}$ is its 4-velocity.
The energy-momentum tensor satisfies the conservation condition $\nabla_{\mu}T^{\mu\nu}=0$,
and its time component is the continuity equation
\begin{alignat}{1}
\dot{\rho}+n H(\rho+p)=0. \label{continue}
\end{alignat}

\section{Equation of the Perfect Fluid in Einstein Gravity}

In Einstein gravity, the field equation is written as
\begin{alignat}{1}
R_{\mu\nu}-\frac{1}{2}g_{\mu\nu}R=8\pi T_{\mu\nu}, \label{fe}
\end{alignat}
and using the metric (\ref{metric}) of the FRW universe and the energy-momentum tensor (\ref{pf}) of the perfect fluid,
one can get the Friedmann's equation
\begin{alignat}{1}
H^2+\frac{k}{a^2}=\frac{16\pi}{n(n-1)}\rho.
\end{alignat}
From the continuity condition (\ref{continue}), one can get
\begin{alignat}{1}
p=-\frac{n-1}{8\pi}\left(\dot{H}-\frac{k}{a^2}\right)-\frac{n(n-1)}{16\pi}\left(H^2+\frac{k}{a^2}\right).
\end{alignat}
Using (\ref{AH}), and (\ref{HT}), one can get the equation of state\footnote{For convenience,
we do not replace $R_A$ with the volume $V$.}
\begin{alignat}{1}
p=-\frac{(n-1)T}{2R_A}-\frac{(n-1)(n-4)}{16\pi R_A^2}. \label{eosEG}
\end{alignat}
The equation of state can be regarded as a function $p=p(R_A,T)$, in order to show its properties, 
one can either keep $R_A$ or $T$ fixed mathematically. However, for dynamical spacetime like the FRW universe, $R_A$ is not fixed, 
so we only fix $T$ and study the behavior of the isothermal lines, which shows the relation between the pressure and the radius of the apparent 
horizon. If $n=3$, for the isothermal line with temperature $T$, the pressure has a minimum $p_{min}=-2\pi T^2$ at $R=1/(4\pi T)$,
and becomes zero at $R=1/(8\pi T)$, see FIG.1.\footnote{Here the temperature $T$ is given as 0.11, 0.1, 0.09 without units, 
which seems strange for some readers. It should be noted that the figures are drawn to show the shape of the isothermal lines and 
the variation trend of the pressure with the temperature, so the specific value or unit is not so important.
It is also a frequently used method in the references about black hole phase transitions, e.g.\cite{Kubiznak:2012wp}.
The values of the temperature in the following figures are chosen on the same spirit.}

\newpage

\begin{figure}[h]
\includegraphics[height=5.5cm]{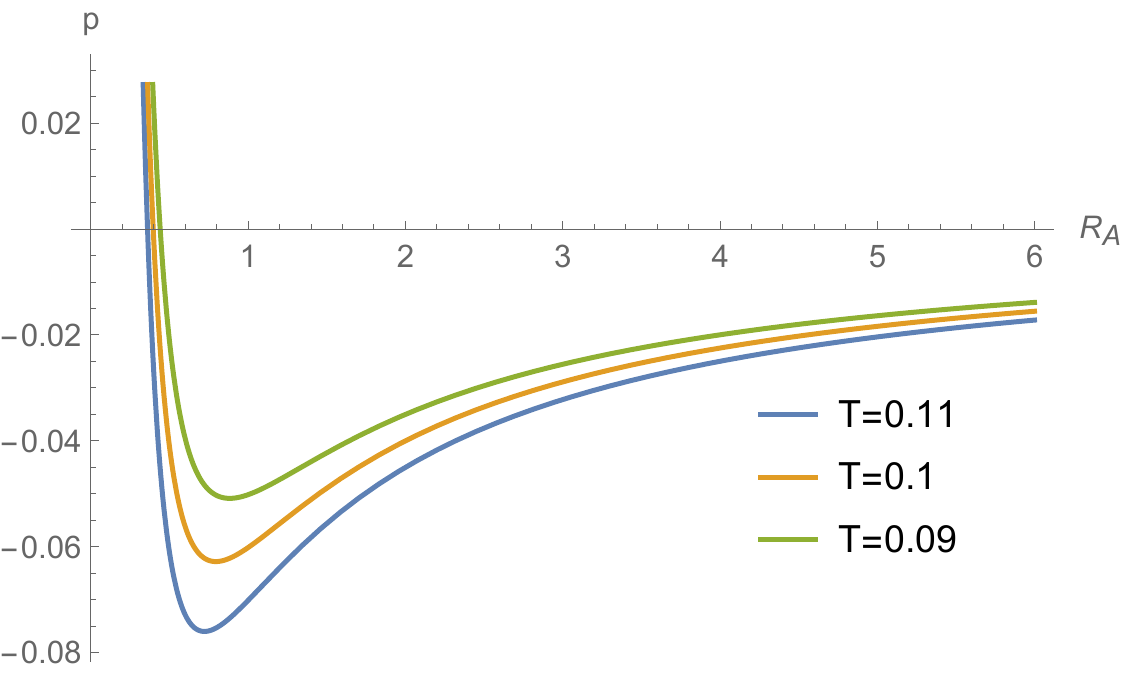}
\caption{Isothermal lines of perfect fluid in the (3+1)-dimensional FRW universe under Einstein gravity.}
\end{figure}

If $n=4$, the equation of state is very simple
\begin{alignat}{1}
p=-\frac{3T}{2R_A},
\end{alignat}
which is always negative, see FIG.2.
\begin{figure}[h]
\includegraphics[height=5.5cm]{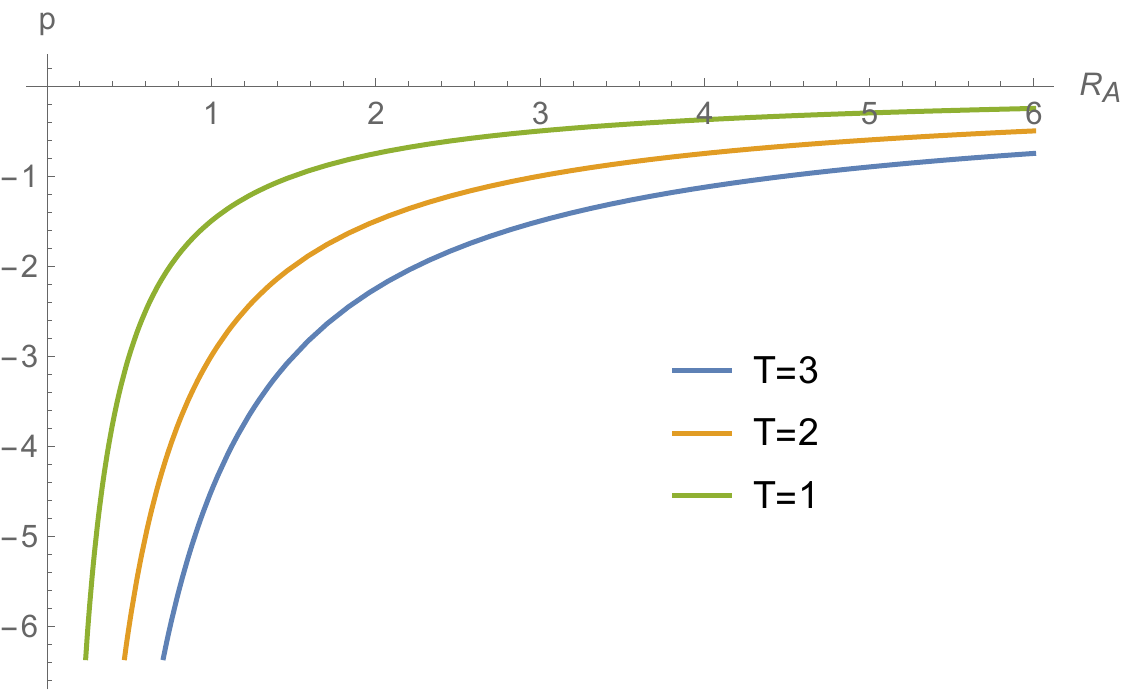}
\caption{Isothermal lines of the perfect fluid in the (4+1)-dimensional FRW universe under Einstein gravity.}
\end{figure}

If $n\geq 5$, the pressure is also always negative, e.g. FIG.3.

\begin{figure}[h]
\includegraphics[height=5.5cm]{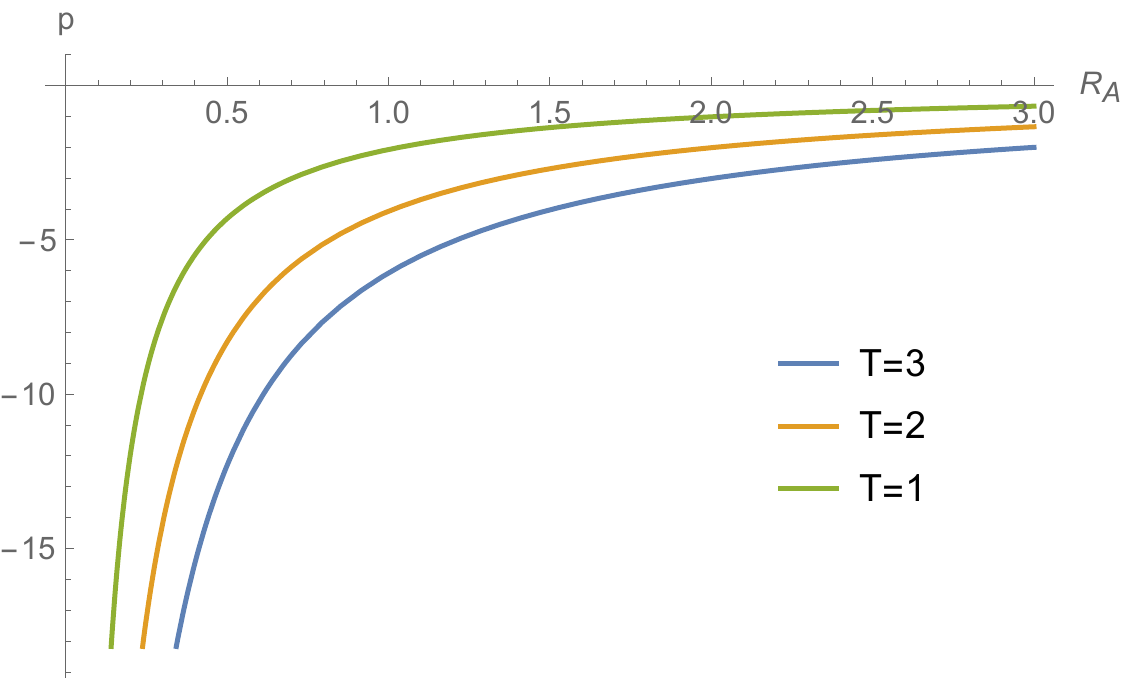}
\caption{Isothermal lines of the perfect fluid in the (5+1)-dimensional FRW universe under Einstein gravity.}
\end{figure}

However, the equation of state (\ref{eosEG}) has no critical point that satisfies the conditions (\ref{CC})
and there is no phase transition in Einstein gravity.

\section{Equation of the Perfect Fluid in Gauss-Bonnet Gravity}

Gauss-Bonnet gravity is one of the most common and simple modified theories of gravity.
It has high-order derivative terms of the curvature that are combined in a certain way, i.e. the Gauss-Bonnet term\cite{Akbar:2006kj},
\begin{alignat}{1}
R_{GB}=R^2-4R_{\mu\nu}R^{\mu\nu}+R_{\mu\nu\rho\sigma}R^{\mu\nu\rho\sigma},
\end{alignat}
which has dynamical effects in dimensions higher than four.
The action of Gauss-Bonnet gravity is written as
\begin{alignat}{1}
S=\frac{1}{16\pi}\int\od^{n+1} x\sqrt{-g}(R+\alpha R_{GB})+S_m,
\end{alignat}
where $S_m$ stands for the action of matter such as the perfect fluid.

The Friedmann's equation in Gauss-Bonnet gravity is
\begin{alignat}{1}
H^2+\frac{k}{a^2}+\tilde{\alpha}\left(H^2+\frac{k}{a^2}\right)^2=\frac{16\pi}{n(n-1)}\rho,
\end{alignat}
where $\tilde{\alpha}=(n-2)(n-3)\alpha$, so it has no difference with Einstein gravity for $n=2,3$.
From the continuity condition (\ref{continue}), one can get
\begin{alignat}{1}
p=&-\frac{n-1}{8\pi}\left(\dot{H}-\frac{k}{a^2}\right)
-\frac{(n-1)\tilde{\alpha}}{4\pi}\left(H^2+\frac{k}{a^2}\right)\left(\dot{H}-\frac{k}{a^2}\right)
\nonumber \\&
-\frac{n(n-1)}{16\pi}\left(H^2+\frac{k}{a^2}\right)-\frac{n(n-1)\tilde{\alpha}}{16\pi}\left(H^2+\frac{k}{a^2}\right)^2,
\end{alignat}
and using (\ref{AH}), and (\ref{HT}), one can get the equation of state
\begin{alignat}{1}
p=-\frac{(n-1)T}{2R_A}-\frac{(n-1)(n-4)}{16\pi R_A^2}-\frac{(n-1)\tilde{\alpha}T}{R_A^3}-\frac{(n-1)(n-8)\tilde{\alpha}}{16\pi R_A^4}.
\label{eosGB}
\end{alignat}

If $n=2,3$, the $\tilde{\alpha}$ terms vanish, so the equation of state is the same as the one from Einstein gravity (\ref{eosEG})
and has no critical point, i.e. (\ref{CC}) have no solutions. If $n=4,5,6,7$ and $\tilde{\alpha}<0$, there exist critical points
satisfying (\ref{CC}), and phase transitions occur above the critical temperature. If $n\geq 8$, there is no critical point.

If $n=4$, the critical point is
\begin{alignat}{1}
R_c=\sqrt{-2\tilde{\alpha}}, \quad T_c=\frac{1}{2\pi\sqrt{-2\tilde{\alpha}}}, \quad p_c=\frac{3}{16\pi\tilde{\alpha}}.
\end{alignat}
Interestingly, the isothermal lines intersect at the critical point as shown in FIG.4.

\begin{figure}[h]
\includegraphics[height=5.5cm]{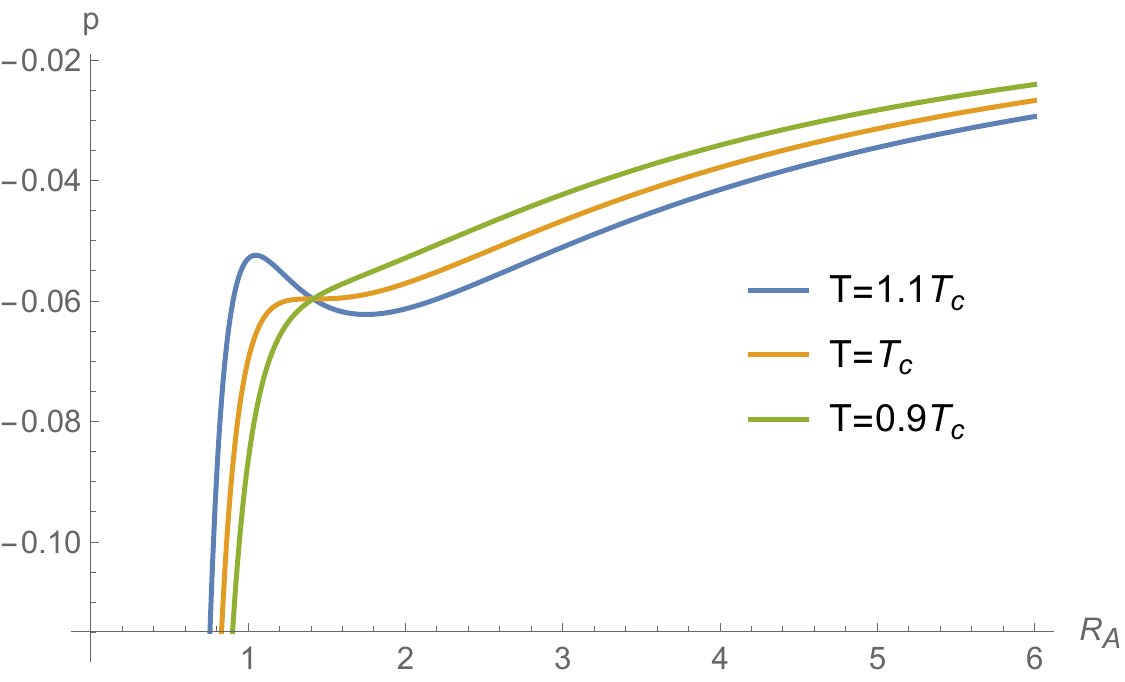}
\caption{Isothermal lines of the perfect fluid in (4+1)-dimensional FRW universe under Gauss-Bonnet gravity with $\tilde{\alpha}=-1$.}
\end{figure}

It can be seen that if the temperature is higher than the critical temperature, the isothermal line has one maximum and 
one minimum, but if the temperature is lower than the critical temperature, the isothermal line is trivial and has neither maxima  
nor minima. The waved line at $T>T_c$ is unstable and should be replaced by a horizontal line according to the Maxwell's equal area law
similar to the $T<T_c$ case in the vdW system.

One can also get a dimensionless quantity
\begin{alignat}{1}
\rho=\frac{p_cv_c}{T_c}=1.
\end{alignat}

If $n=5$ and $\tilde{\alpha}<0$, the critical point is
\begin{alignat}{1}
R_c=\sqrt{6(2-\sqrt{5})\tilde{\alpha}}, \quad T_c=\frac{(7+3\sqrt{5})\sqrt{6(\sqrt{5}-2)}}{48\pi\sqrt{-\tilde{\alpha}}},
\quad p_c=\frac{23+10\sqrt{5}}{144\pi\tilde{\alpha}},
\end{alignat}
and
\begin{alignat}{1}
\rho=\frac{11+\sqrt{5}}{18}\approx 0.735,
\end{alignat}
see FIG.5.
\begin{figure}[h]
\includegraphics[height=5.5cm]{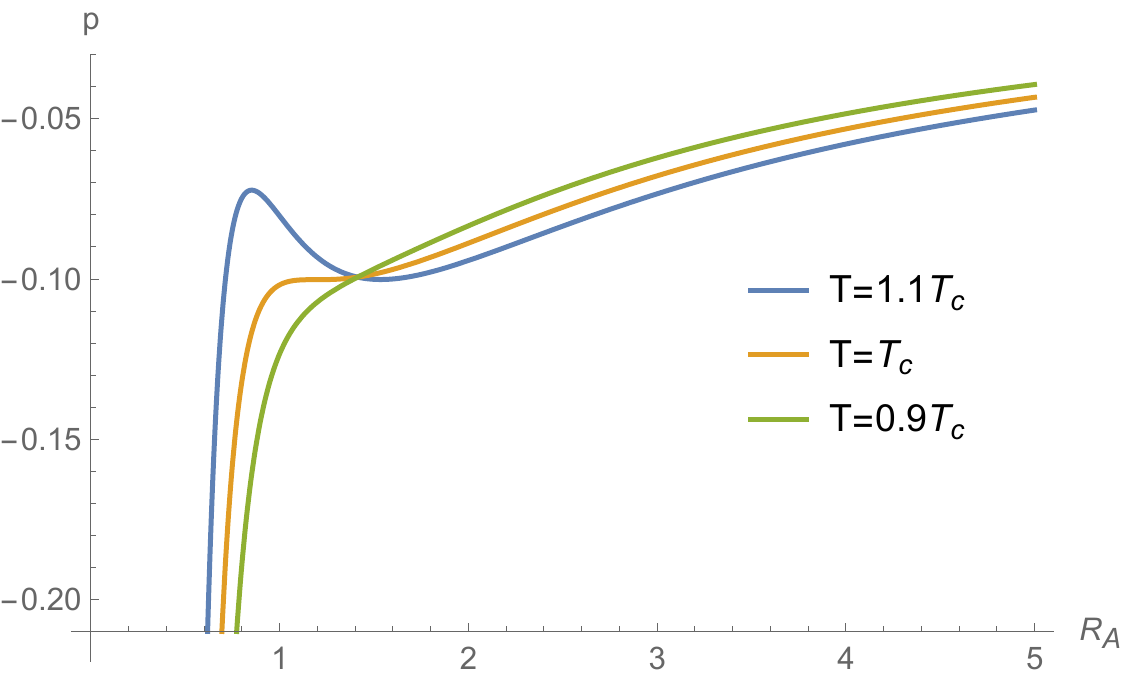}
\caption{Isothermal lines of the (5+1)-dimensional FRW universe in Gauss-Bonnet gravity with $\tilde{\alpha}=-1$.}
\end{figure}

If $n=6$ and $\tilde{\alpha}<0$, the critical point is
\begin{alignat}{1}
R_c=\sqrt{2(3-2\sqrt{3})\tilde{\alpha}}, \quad T_c=\frac{(2+\sqrt{3})\sqrt{2(2\sqrt{3}-3)}}{12\pi\sqrt{-\tilde{\alpha}}},
\quad p_c=\frac{5(15+8\sqrt{3})}{288\pi\tilde{\alpha}},
\end{alignat}
and
\begin{alignat}{1}
\rho=\frac{6+\sqrt{3}}{12}\approx 0.644,
\end{alignat}
see FIG.6.

\newpage

\begin{figure}[h]
\includegraphics[height=5.5cm]{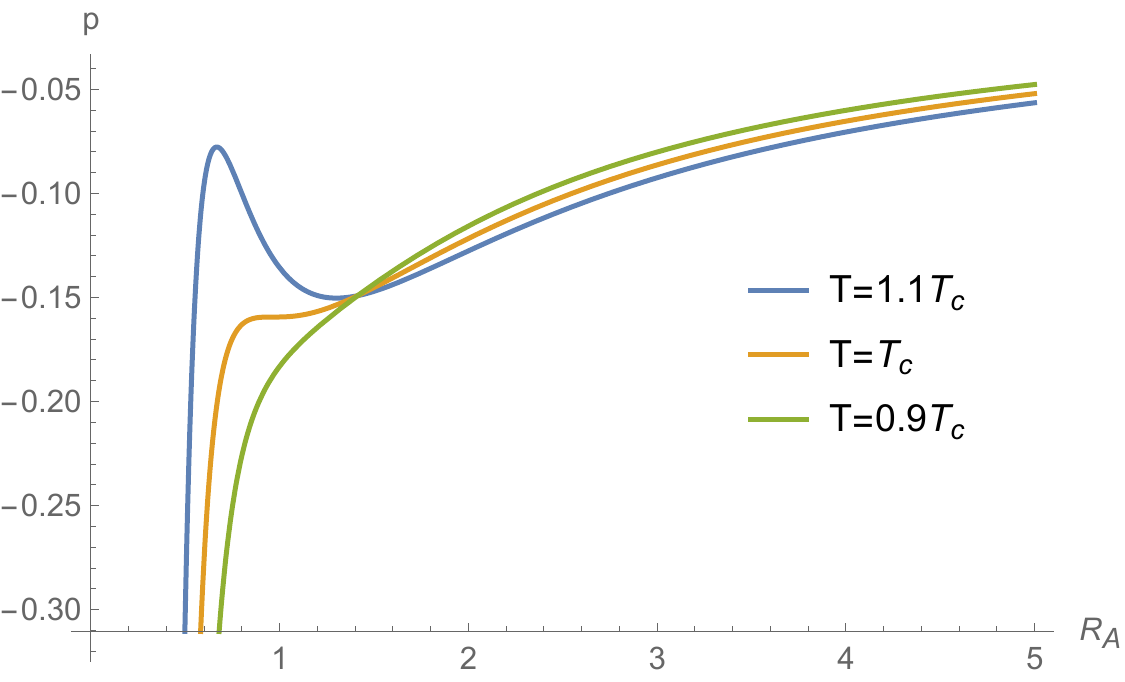}
\caption{Isothermal lines of the perfect fluid in (6+1)-dimensional FRW universe under Gauss-Bonnet gravity with $\tilde{\alpha}=-1$.}
\end{figure}

If $n=7$ and $\tilde{\alpha}<0$, there is also a critical point
\begin{alignat}{1}
R_c=\sqrt{2(2-\sqrt{5})\tilde{\alpha}}, \quad T_c=\frac{(3+\sqrt{5})\sqrt{2(\sqrt{5}-2)}}{16\pi\sqrt{-\tilde{\alpha}}},
\quad p_c=\frac{3(5+2\sqrt{5})}{32\pi\tilde{\alpha}},
\end{alignat}
and
\begin{alignat}{1}
\rho=\frac{5+\sqrt{5}}{8}\approx 0.905,
\end{alignat}
see FIG.7.
\begin{figure}[h]
\includegraphics[height=5.5cm]{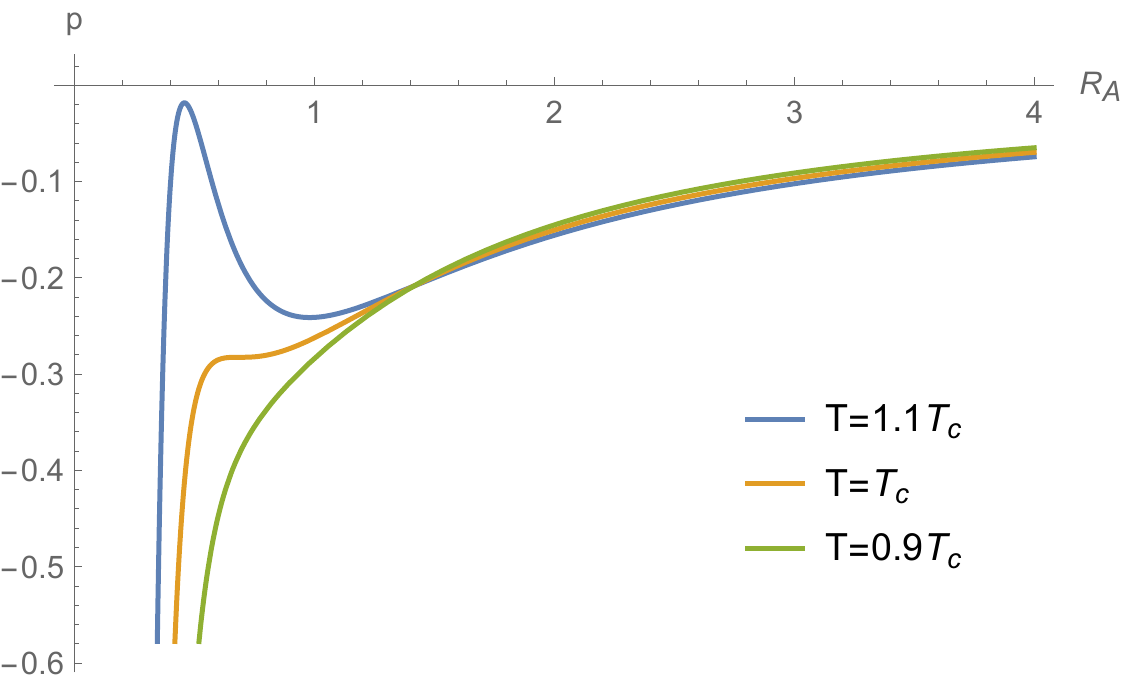}
\caption{Isothermal lines of the (7+1)-dimensional FRW universe in Gauss-Bonnet gravity with $\tilde{\alpha}=-1$.}
\end{figure}

\section{Equation of the Perfect Fluid in Lovelock Gravity}

The Lovelock gravity is a generalization of Einstein gravity and Gauss-Bonnet gravity.
The action in Lovelock gravity can be written as\cite{Cai:2006rs}
\begin{alignat}{1}
S=\frac{1}{16\pi}\int\od^{n+1}x\sqrt{-g}\sum_{i=0}^{m}c_i L_i+S_m,
\end{alignat}
where $c_i$ is an arbitrary constant, $m\leq[n/2]$, $S_m$ still stands for the action of matter,
and $L_i$ is the Euler density of a $2i$-dimensional manifold
\begin{alignat}{1}
L_i=\frac{1}{2^i}\delta^{a_1b_1...a_ib_i}_{c_1d_1...c_id_i}R^{c_1d_1}_{~~~~a_1b_1}...R^{c_id_i}_{~~~~a_ib_i}.
\end{alignat}
The $i=0$ term is equivalent to the cosmological constant, the $i=1$ term is the Einstein-Hilbert term, and the $i=2$ term corresponds to
the Gauss-Bonnet term. From the action, one can get the field equations
\begin{alignat}{1}
\sum_{i=0}^{m}\frac{c_i}{2^{i+1}}\delta^{aa_1b_1...a_ib_i}_{bc_1d_1...c_id_i}R^{c_1d_1}_{~~~~a_1b_1}...R^{c_id_i}_{~~~~a_ib_i}=8\pi T^a_b,
\end{alignat}
where $T^a_b$ is the energy-momentum tensor of the perfect fluid (\ref{pf}).

From the field equations, one can get the Friedmann equation
\begin{alignat}{1}
\sum_{i=0}^{m}\hat{c}_i\left(H^2+\frac{k}{a^2}\right)^{i}\equiv H^2+\frac{k}{a^2}+\sum_{i=2}^{m}\hat{c}_i\left(H^2+\frac{k}{a^2}\right)^{i}
=\frac{16\pi}{n(n-1)}\rho,
\end{alignat}
where we set $\hat{c}_0=0$, $\hat{c}_1=1$, and $\hat{c}_i=c_i n!/(n-2i)!$.
Using the continuity condition (\ref{continue}), one can get
\begin{alignat}{1}
p=&-\frac{n-1}{8\pi}\left(\dot{H}-\frac{k}{a^2}\right)
-\frac{n(n-1)}{16\pi}\sum_{i=2}^{m}\hat{c}_i\left(H^2+\frac{k}{a^2}\right)^{i}
-\frac{n(n-1)}{16\pi}\left(H^2+\frac{k}{a^2}\right)
\nonumber \\ &
-\frac{(n-1)}{8\pi}\sum_{i=2}^{m}i\hat{c}_i\left(H^2+\frac{k}{a^2}\right)^{i-1}\left(\dot{H}-\frac{k}{a^2}\right),
\end{alignat}
and using (\ref{AH}), (\ref{HT}), one can get the equation of state
\begin{alignat}{1}
p=-\frac{(n-1)T}{2R_A}-\frac{(n-1)(n-4)}{16\pi R_A^2}-\frac{n-1}{2}\sum_{i=2}^{m}\frac{i\hat{c}_i T}{R_A^{2i-1}}
+\frac{n-1}{16\pi}\sum_{i=2}^{m}\frac{(4i-n)\hat{c}_i}{R_A^{2i}}. \label{eosLove}
\end{alignat}
Applying the critical conditions (\ref{CC}), one can get
\begin{alignat}{1}
\frac{T}{2R_A^2}+\frac{n-4}{8\pi R_A^3}+\sum_{i=2}^{m}\frac{i(2i-1)\hat{c}_i T}{2R_A^{2i-2}}
-\sum_{i=2}^{m}\frac{i(4i-n)\hat{c}_i}{8\pi R_A^{2i-1}}=&0, \label{cc1}
\\
\frac{T}{R_A^3}+\frac{3(n-4)}{8\pi R_A^4}+\sum_{i=2}^{m}\frac{i(i-1)(2i-1)\hat{c}_i T}{R_A^{2i-1}}
-\sum_{i=2}^{m}\frac{i(2i-1)(4i-n)\hat{c}_i}{8\pi R_A^{2i}}=&0. \label{cc2}
\end{alignat}
From the above two equations, one can get the condition satisfied by the critical radius 
\begin{alignat}{1}
n-4+\sum_{i=0}^{2m-4}\frac{C_i}{R_A^{i}}+\sum_{i,j=2}^{m}\frac{C_{ij}}{R_A^{2i+2j}}=0,
\end{alignat}
where $C_i, C_{ij}$ are some coefficients, and its solution can be written as $R_c$.
Insert $R_c$ into (\ref{cc1}) or (\ref{cc2}), one can get the critical temperature $T_c$,
and insert $R_c$ and $T_c$ into (\ref{eosLove}), one can get the critical pressure $P_c$.

\section{Conclusions and Discussions}

In this paper, we obtain the van der Waals-like equation for the perfect fluid in the $D$-dimensional FRW universe under Einstein gravity, Gauss-Bonnet gravity and Lovelock gravity respectively. In Einstein gravity, the isothermal lines of the perfect fluid in the 4-dimensional
FRW universe have minima but no critical point, and the isothermal lines of the perfect fluid in higher-dimensional FRW universe have neither minimum nor critical point. In Gauss-Bonnet gravity with $\alpha<0$, the equations of the perfect fluid in the $5,6,7,8$-dimensional FRW universe have critical points and phase transitions occur above the critical temperature. In Lovelock gravity, we also get the equation of the perfect fluid and the condition of the critical points in the $D$-dimensional FRW universe.

The results show that critical points and phase transitions of the perfect fluid depend on both the theories of gravity and the dimensions of the FRW universe.
If the critical point exists, phase transitions always take place above the critical temperature, which appears to be a universal feature.

\section*{Acknowledgments}

We thank Haximjan Abdusattar for the discussions. This work is supported by National Natural Science Foundation of China (NSFC) under grant No.12465011
and East China University of Technology (ECUT) under grant DHBK2023002.

\end{document}